\documentclass[prl,aps,showpacs,twocolumn,floatfix]{revtex4-1}
\usepackage{amsmath,amssymb,graphicx,cases}
\usepackage{epsfig}
\usepackage{textcomp}
\usepackage{epstopdf}
\usepackage{float}
\usepackage[normalem]{ulem}
\newcommand{\stkout}[1]{\ifmmode\text{\sout{\ensuremath{#1}}}\else\sout{#1}\fi}

\def\blue{\textcolor{blue}}

\usepackage[colorlinks=true, urlcolor=blue, anchorcolor=blue, citecolor=blue,filecolor=blue,linkcolor=blue,menucolor=blue,pagecolor=blue]{hyperref}

\begin{document}
\title{Narrow Escape of Interacting Diffusing Particles}
\author{Tal Agranov}
\email{tal.agranov@mail.huji.ac.il}
\affiliation{Racah Institute of Physics, Hebrew University of
Jerusalem, Jerusalem 91904, Israel}
\author{Baruch Meerson}
\email{meerson@mail.huji.ac.il}
\affiliation{Racah Institute of Physics, Hebrew University of
Jerusalem, Jerusalem 91904, Israel}


\begin{abstract}
The narrow escape problem deals with the calculation of the mean escape time (MET) of a Brownian particle from a bounded domain through a small hole on the domain's boundary.  Here we develop a formalism that allows us to evaluate the \emph{non-escape probability} of a gas of diffusing particles that may interact with each other. In some cases the non-escape probability allows us to evaluate the MET of the first particle. The formalism is based on the fluctuating hydrodynamics and the recently developed macroscopic fluctuation theory. We also uncover an unexpected connection between the narrow escape of interacting particles and thermal runaway in chemical reactors.
\end{abstract}
\maketitle

\begin{figure} [ht]
	\includegraphics[width=0.40\textwidth,clip=]{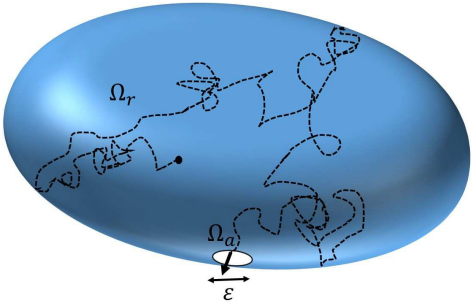}
	\caption{Narrow escape of a single Brownian particle.  $\Omega_{a}$ is a small hole of linear size  $\epsilon$.  $\Omega_r$ is the reflecting part of the boundary.}
	\label{hure}
\end{figure}

The narrow escape problem (NEP) \cite{keller1,bress,beni,bookz,rev,russians} is ubiquitous in physics, chemistry, and biology. It deals with the calculation of the mean time it takes a Brownian particle inside a bounded domain to escape through a narrow window on the domain's boundary, see Fig. \ref{hure}.
In the past two decades this beautiful and mathematically intricate problem has received much attention, as it was realized that the mean escape time (MET) controls the rates of many important processes in molecular and cellular biology, such as
arrival of a receptor at a reaction site on the surface of a cell \cite{diffapfere}, transport of RNA molecules from the nucleus to the cytoplasm through nuclear pores \cite{rna}, diffusion of calcium ions in dendritic spines \cite{den}, and other processes \cite{bress}.
When the size of the escape hole $\epsilon$ is much smaller than the domain size $L$, the MET of a Brownian particle can be expressed via the principal eigenvalue of the Laplace's operator inside the domain with the absorbing (Dirichlet) boundary condition on the escape hole and the reflecting (Neumann) condition on the rest of the boundary, see \textit{e.g.} \cite{russians}. The latter problem goes back to Helmholtz \cite{helmholtz} and Lord Rayleigh \cite{rayleigh}. Recent theoretical developments addressed the role of the initial position of the Brownian particle \cite{beni2}, complicated geometries \cite{geo1,geo2,geo3,geo4,geo5,geo6,chevi1,chevi2,Greb2016}, finite lifetime of the escaping particle \cite{kill,kill2}, and the presence of a kinetic bottleneck at the escape hole \cite{gleb}.

In a host of situations of biological importance there are \emph{many} Brownian particles, which attempt to escape through a small hole (or reach a small site). If they are treated as non-interacting, the escape statistics
can be expressed via the one-particle statistics \cite{RoKim,many2}. Quite often, however, the particles interact with each other, such as in a highly crowded intracellular environment \cite{bress}. Although the importance of interactions
may have been recognized earlier, there have been no attempts to include them
in the theory. This is our main objective here, but the formalism proves useful also for ensembles of non-interacting particles.

One approach to solving the NEP for a \emph{single} Brownian particle with diffusivity $D_0$ relies on the calculation of the particle's non-escape probability until time $T$, $\mathcal P_1(T)$. In the small-window limit, $\epsilon/L\ll 1$, the problem simplifies because the particle's escape becomes a relatively rare event \cite{russians}. For times much longer than the diffusion time across the escape hole, $T\gg \epsilon^2/D_0$, and for a uniformly distributed random initial position of the particle, $\mathcal P_1(T)$ decays exponentially in time \cite{keller1,bress,beni,bookz,russians},
\begin{equation}\label{suas}
-\ln\mathcal P_1\left(T\right)\simeq TD_0 \mu_0^2 ,
\end{equation}
where $\mu_0^2$ is the principal eigenvalue of the eigenvalue problem $\nabla^2 \Psi+\mu^2 \Psi=0$ inside the domain with the mixed boundary conditions $\Psi(\mathbf{x} \in  \Omega_a,t) = \nabla\Psi(\mathbf{x} \in \Omega_r,t)\cdot\hat{n}=0$. Here $\Omega_r$ is the reflecting part of the domain's boundary, $\Omega_a$ is the complementary absorbing part (the small escape hole), and $\hat{n}$ is the local normal to the boundary.
Correspondingly, the MET is equal to $\langle T_1 \rangle\simeq 1/ D_0\mu_0^2$, and this result holds up to small corrections in $\epsilon/L$ \cite{chevi1}.
In the leading  order,  which is $O(\epsilon/L)$, $\mu_0^2$ (found already by Lord Rayleigh \cite{rayleigh}) can be expressed through the electrical capacitance $C_{\epsilon}$ of the \emph{conducting} patch $\Omega_a$ in an otherwise empty space: $\mu_0^2\simeq 2\pi C_{\epsilon}/V$, where $V$ is the domain's volume. The capacitance  $C_{\epsilon}$ scales as $\epsilon$. If $\Omega_a$ is a disk  of radius $\epsilon$, then $C_{\epsilon}=2\epsilon/\pi$ \cite{jackson} leading to $\langle T_1 \rangle\simeq  V/(4\epsilon D_0)$, which is independent of the domain shape \cite{russians}.

The non-escape probability $\mathcal{P}(T,N)$ of $N$ non-interacting Brownian particles, randomly distributed over the domain, is the product of their single-particle non-escape probabilities (\ref{suas}). Therefore, at long times, it also decays exponentially in time,
$-\ln{\mathcal P}(T,N)\simeq T s(n_0,\epsilon)$, with the decay rate
\begin{equation}\label{asuas}
s(n_0,\epsilon)= N D_0\mu_0^2\simeq 2\pi C_{\epsilon}  D_0n_0,
\end{equation}
where $n_0=N/V$ is the particle number density. For very low densities, $n_0\epsilon^3\ll1$, Eq.~(\ref{asuas}) yields the MET of the first particle, $\langle T \rangle\simeq1/s(n_0,\epsilon)$ \cite{RoKim}. Indeed, in this regime $\langle T \rangle$ is much longer than the diffusion time across the hole, $\epsilon^2/D_0$.

At higher densities,  $n_0\epsilon^3\gg 1$, we have $\langle T \rangle\ll \epsilon^2/D_0$. As the diffusion length scale $\sqrt{D_0 T}$ is now much smaller than $\epsilon$, the process is effectively one-dimensional in the direction normal to the hole. Here the non-escape problem reduces to a well-studied problem of finding the survival probability $\mathcal P_{\text{1d}}$ of a gas of non-interacting Brownian particles of density $n_{1d}$ (per unit length), randomly placed  on a half-line $x>0$, against absorption at $x=0$ \cite{1d1,1d2,1d3,1d4,1d5,1d6,1d7,1d8,1d9,MVK}. Here $\mathcal P_{\text{1d}}$ decays as a stretched exponential, $-\ln {\mathcal P_{1d}} \simeq (2/\sqrt{\pi})  n_{1d} \sqrt{D_0T}$ \cite{1d2,MVK}. To evaluate ${\mathcal P}(T,N)$, one should set $n_{1d}=n_0 A_{\epsilon}$, where $A_{\epsilon}$ is
the area of $\Omega_{\epsilon}$ \cite{MVK}.  For a circular hole of radius $\epsilon$ this leads to
\begin{equation}\label{shorttimeRW}
-\ln {\mathcal P}(T,N)\simeq 2 \sqrt{\pi} n_0 \epsilon^2 \sqrt{D_0 T} ,
\end{equation}
and one obtains $\langle T \rangle\simeq (2\pi  D_0 n_0^2 \epsilon^4)^{-1}$ \cite{RoKim}.

For interacting particles the non-escape probability $\mathcal{P}(T,N)$ is \emph{not} equal to the product of single-particle probabilities, and a new approach is required. We develop such an approach here and calculate the non-escape probability $\mathcal{P}(T,N)$ of $N\gg 1$ interacting particles at long and short times. At long times, $\mathcal{P}(T,N)$ decays exponentially in time,
\begin{equation}\label{non-escape}
-\ln\mathcal P(T,N)\simeq Ts(n_0,\epsilon).
\end{equation}
The dependence of $s(n_0,\epsilon)$ on the geometry factorizes up to small corrections in $\epsilon/L$. In the leading order in $\epsilon/L$ we obtain
\begin{equation}
s\left(n_0,\epsilon\right) \simeq \pi  C_{\epsilon} f^2(n_0).
\label{sout}
\end{equation}
The nonlinear function $f(n_0)$, which we show how to calculate,
encodes particle interactions and is model-dependent.

At short times we obtain
\begin{equation}
-\ln {\mathcal P}(T,N)\simeq A_{\epsilon} g(n_0)  \sqrt{D_0T},
\label{shorttimeint}
\end{equation}
with a model-dependent nonlinear function $g(n_0)$.

Now we present our results in some detail. Assuming a large number of particles in the relevant regions of space, we employ \emph{fluctuating hydrodynamics}: a coarse-grained description in terms of the (fluctuating) particle number density $\rho(\mathbf{x},t)$ \cite{Spohn,KL}. The \emph{average} particle density obeys a diffusion equation $\partial_t \rho = \nabla \cdot \left[D(\rho) \nabla \rho\right]$, whereas macroscopic fluctuations are described by the conservative Langevin equation
\begin{equation}
\partial_t \rho = -\nabla \cdot \mathbf{J} ,\quad\mathbf{J}=-D(\rho) \nabla \rho-\sqrt{\sigma(\rho)}\boldsymbol{\eta}(\mathbf{x},t),\label{lang}
\end{equation}	
where $D(\rho)$ and $\sigma(\rho)$ are the diffusivity and mobility of the gas of particles, and $\boldsymbol{\eta}(\mathbf{x},t)$ is a zero-mean Gaussian noise, delta-correlated in space and time. The density $\rho$ and flux $\mathbf{J}$ satisfy the boundary conditions
\begin{equation}\label{blang}
\rho(\mathbf{x}\in{\Omega_a},t)=0, \quad\mathbf{J}(\mathbf{x}\in{\Omega_r},t)\cdot\hat{n}=0.
\end{equation}

To proceed further we employ the recently developed \emph{macroscopic fluctuation theory} (MFT) \cite{MFTreview}. The MFT grew from the Martin-Siggia-Rose path integral formalism in physics \cite{MSR,deridamft,map} and the Freidlin-Wentzell large-deviation theory in mathematics \cite{Freid}. It follows from a path integral formulation for Eq.~(\ref{lang}), which describes the probability of observing a joint density and flux histories $\rho(\mathbf{x},t), \mathbf{J}(\mathbf{x},t)$, constrained by the conservation law (\ref{lang}),
\begin{eqnarray}\nonumber
&&\!\!\!\!\!\!\mathcal P =\int\mathcal{D}\rho\mathcal{D}\mathbf{J}
\prod_{\mathbf{x},t}\delta(\partial_{t}{\rho}+\nabla \cdot \mathbf{J})\,
 \exp\left(-\mathcal{S}\right),\label{path0}\\
&&\!\!\!\!\!\!\mathcal{S}\left[\rho(\mathbf{x},t),\mathbf{J}(\mathbf{x},t)\right]=\int_0^Tdt\int d^3\mathbf{x}\frac{\left[\mathbf{J}+D(\rho)\nabla \rho\right]^2}{2\sigma(\rho)}.\label{path1}
\end{eqnarray}
The next step in the derivation, by now fairly standard \cite{MFTreview,map,deridamft}, exploits the large parameter $N\gg 1$ to perform a saddle-point evaluation of the path integral. The dominant contribution to $\mathcal P$ comes from the \emph{optimal fluctuation}: the most probable history $(\rho,\mathbf{J})$ ensuring the particle non-escape up to the specified time $T$
and obeying the conservation law. The ensuing minimization procedure yields the Euler-Lagrange equation and the problem-specific boundary conditions. With the solutions at hand, one
calculates the action $S$, which yields the non-escape probability ${\mathcal P}(T,N)$ up to a pre-exponential factor,
\begin{eqnarray}
\label{actionmain}
-\ln {\mathcal P}(T,N)
\simeq S \equiv\min_{\rho,\mathbf{J}}\mathcal{ S}\left[\rho(\mathbf{x},t),\mathbf{J}(\mathbf{x},t)\right].
\end{eqnarray}

The resulting problem simplifies in the limits of very long and very short times (we elaborate on the relevant time scales below). At long times, the optimal gas density and flux, conditioned on non-escape, become \emph{stationary}, in analogy with a  closely related problem  of survival of particles inside domains with \emph{fully} absorbing boundaries \cite{surv}. As a result, ${\mathcal P}(T,N)$  exponentially decays with time $T$, see Eq.~(\ref{non-escape}). A similar property lies at the origin of the ``additivity principle"  \cite{bd}, proposed in the context of \emph{stationary} fluctuations of current in systems driven by density reservoirs at the boundaries.

In the stationary formulation, Eq.~(\ref{lang}) yields $\nabla \cdot \mathbf{J} = 0$, so the optimal flux $\mathbf{J}$ is a solenoidal vector field. In the non-escape problem, $\mathbf{J}$ must also have zero normal component at the \emph{entire} domain's boundary.
Using these properties, one can show (see Ref. \cite{surv} and Appendix A of Ref. \cite{void}) that $\mathbf{J}$ is also vortex-free and thus vanishes identically. This means that the
fluctuating contribution to the optimal flux exactly counterbalances the deterministic contribution, thus preventing the particles from escaping. Now we have to find the optimal  density profile. Upon the ansatz $\mathbf{J}=0$ and $\rho=\rho(\mathbf{x})$ in Eq.~(\ref{path1}), the action $\mathcal{S}$ becomes proportional to $T$, and the problem reduces to minimizing the \emph{action rate} functional
\begin{equation}\label{lag}
\mathfrak{s}\left[\rho\left(\mathbf{x}\right)\right]=\int d^3\mathbf{x}\frac{\left[D(\rho)\nabla \rho\right]^2}{2\sigma(\rho)},
\end{equation}
subject to the boundary conditions (\ref{blang}) and the mass conservation constraint
\begin{equation}\label{cons}
\int d^3\mathbf{x}\,\rho(\mathbf{x})=n_0 V.
\end{equation}
Let us introduce the new variable $u(\mathbf{x})=f\left[\rho\left(\mathbf{x}\right)\right]$,
where the function $f$ is defined by the integral \cite{conv}:
\begin{equation}
f(\rho)=\int_0^\rho dz \frac{D(z)}{\sqrt{\sigma (z)}}.\label{transform1}
\end{equation}
We denote the inverse function, $f^{-1}$, by $F$. Expressed through $u(\mathbf{x})$, the action rate~(\ref{lag}) is reduced to the effective ``electrostatic action"
\begin{equation}
\label{lang2}
\mathfrak{s}\left[u\left(\mathbf{x}\right)\right]=\frac{1}{2}\int d^3\mathbf{x}\left[\nabla u(\mathbf{x})\right]^2,
\end{equation}
which, remarkably, is universal for all interacting particle models described by Eq.~(\ref{lang}).
Now we minimize this action, incorporating the mass conservation (\ref{cons}),
\begin{equation}\label{cons2}
\int d^3\mathbf{x} \,F\left[u\left(\mathbf{x}\right)\right]=n_0 V,
\end{equation}
via a Lagrange multiplier $\Lambda$. The Euler-Lagrange equation has the form of
a non-linear Poisson equation \cite{surv},
\begin{equation}
\nabla^2 u +  \Lambda \frac{d F(u)}{du}=0, \label{ustatnd}
\end{equation}
with the mixed boundary conditions \cite{boundary},
\begin{equation}
u(\mathbf{x}\in{\Omega_a})=0 ,\quad\nabla u(\mathbf{x}\in{\Omega_r})\cdot\hat{n}=0.
\label{bqu1}
\end{equation}
The action rate~(\ref{lang2}), evaluated on the solution to the problem (\ref{cons2})--(\ref{bqu1}), yields the decay rate $s(n_0,\epsilon)$ from Eq.~(\ref{non-escape}), specific to each gas model. If there are
multiple solutions,
the minimum-action solution must be chosen.

Now we apply the steady-state formalism to the \emph{diffusive lattice gases }\cite{Spohn,KL,Liggett}. This is a class of microscopic models, defined by a prescribed stochastic particle dynamics on a lattice.
The diffusivity $D(\rho)\geq 0$ and the mobility $\sigma(\rho)\geq 0$ should be obtained from the microscopic model. The simplest example is a gas of non-interacting random walkers (RWs). On large scales and at long times
these are indistinguishable from the non-interacting Brownian particles \cite{Paulbook}. For the RWs one has $D(\rho)=D_0=\text{const}$, and $\sigma(\rho)=2D_0\rho$ \cite{Spohn}.

A more interesting example is the symmetric simple exclusion process (SSEP),
which accounts for excluded-volume interactions. In the SSEP each particle can hop to a neighboring lattice site only if that site is vacant \cite{Spohn}. In the coarse-grained description of the SSEP one has $D(q)=D_0=\text{const}$ and $\sigma(\rho)=2D_0\rho(1-\rho a^3)$ \cite{Spohn,KL}. We set the lattice constant $a$ to unity, so that  $0\leq\rho\leq1$.

Let us first see that the formalism (\ref{transform1})--(\ref{bqu1}) reproduces the classical narrow-escape results for the RWs. In this case Eq.~(\ref{transform1}) yields $f(\rho)=\sqrt{2D_0\rho}$, while Eq.~(\ref{ustatnd}) reduces to the Helmholtz equation
\begin{equation}\label{helm2}
\nabla^2 u+\mu^2 \,u=0 ,
\end{equation}
with $ \mu^2\equiv\Lambda/D_0$ playing the role of the eigenvalue. The minimum action is achieved for the fundamental mode of this equation. We denote it by $\Psi_0(\mathbf{x})$ and normalize
it to unity, $\int d^3\mathbf{x}\,\Psi_0^2(\mathbf{x}) =1$.
Subject to the mass conservation (\ref{cons2}), the solution can be written as $u\left(\mathbf{x}\right)=\sqrt{2ND_0}\Psi_0(\mathbf{x})$. Now we plug it into Eq.~(\ref{lang2}),
use the identity $(\nabla \Psi_0)^2  = \nabla\cdot (\Psi_0 \nabla \Psi_0)-\Psi_0 \nabla^2 \Psi_0$,
apply the divergence theorem to the first term on the right, and use Eqs.~(\ref{bqu1}) and~(\ref{helm2}) for $\Psi_0(\mathbf{x})$. The resulting $\mathfrak{s}[u(\mathbf{x})]= s(n_0,\epsilon)$ is equal to $N D_0\mu_0^2$ in agreement with the exact result cited in Eq.~(\ref{asuas}). The case of RWs is important because here one can also exactly solve the full \emph{time-dependent} MFT equations \cite{surv}. The time-dependent solution shows that, for $T \gg\epsilon^2/D_0$, the leading-order contribution to the action indeed comes from
the steady-state solution. Furthermore, only a vicinity of the escape hole contributes. That is, to leading order in $\epsilon/L$, the solution for a finite domain coincides with the one for
a gas of particles occupying the infinite half-space on one side of an infinite reflecting plane with the hole $\Omega_a$ on it.

For interacting particles Eq.~(\ref{ustatnd}) is nonlinear, but we can exploit the small
parameter $\epsilon/L$ in the same spirit.  The non-escape probability of the gas in the infinite half-space until a long time $T$ can be obtained from an \emph{unconstrained} minimization procedure where, instead of Eq.~(\ref{cons2}), we use the boundary condition $u(\mathbf{x}\rightarrow\infty)=f(n_0)$. Setting $\Lambda=0$ in Eq.~(\ref{ustatnd}), we arrive at the Laplace's equation for $u(\mathbf{x})$. The solution can be expressed through the electrostatic potential $\phi(\mathbf{x})$  of a conducting patch $\Omega_a$ kept at unit voltage on an otherwise insulating infinite plane,
\begin{equation}
u(\mathbf{x})=f(n_0)\left[1-\phi(\mathbf{x})\right].\label{u0}
\end{equation}
In simple cases (\textit{e.g.}, when $\Omega_a$ is a disk), $\phi(\mathbf{x})$ can be found explicitly \cite{jackson}.
Equation~(\ref{u0}) yields the stationary density profile, optimal for the particle non-escape: $\rho(\mathbf{x})=F\{f(n_0)[1-\phi(\mathbf{x})]\}$.
Plugging Eq.~(\ref{u0}) in Eq.~(\ref{lang2}) yields the announced result (\ref{sout}) for the decay rate of the non-escape probability to order $\epsilon/L$. It is given by the electrostatic energy created by a conductor $\Omega_a$ held at voltage $f(n_0)$,
where $C_{\epsilon}$ is the electrical capacitance of the conductor $\Omega_a$. The entire effect of interactions is encoded in
the density dependence $f(n_0)$, coming from the nonlinear transformation~(\ref{transform1}). The geometry dependence is universal for all gases of this class and is given by the capacitance $C_{\epsilon}$. The latter is determined by the shape of the hole and is independent of the domain shape. A dependence on the domain shape emerges in higher orders in $\epsilon/L$. When specialized to the RWs, Eq.~(\ref{sout}) yields  the approximate result cited in Eq.~(\ref{asuas}), as to be expected.

\begin{figure}
\includegraphics[width=0.30\textwidth,clip=]{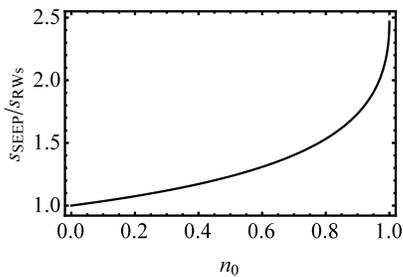}
\caption{The ratio of the decay rate of the non-escape probability for the SSEP, Eq.~(\ref{soutssep}), to the same quantity for the RWs, Eq.~(\ref{asuas}), vs the gas density $n_0=N/V$.}
\label{t}
\end{figure}

For the SSEP  Eq.~(\ref{transform1}) yields $f(\rho)=\sqrt{2D_0}\arcsin(\sqrt{\rho})$, whereas for a small circular window
of radius $\epsilon$ we have $C_{\epsilon}=2\epsilon/\pi$. The resulting decay rate of $\mathcal{P}(T,N)$  is
\begin{equation}
s(n_0,\epsilon)\simeq 4D_0\epsilon \arcsin^2(\sqrt{n_0}).\label{soutssep}
\end{equation}
Figure \ref{t} shows the density dependence of the \emph{ratio} of this decay rate to the decay rate for the RWs, Eq.~(\ref{asuas}). At finite densities
this ratio is always larger than $1$, as to be expected because of the effective
mutual repulsion of the SSEP particles. The finite value of the ratio, $\pi^2/4$, at close packing of the SSEP  should not be taken too seriously, because
fluctuating hydrodynamics breaks down here \cite{surv}. For low densities $n_0\epsilon^3<<1$, the MET of the first particle is given by $\langle T \rangle \simeq 1/s(n_0,\epsilon)$.

Higher-order corrections (with respect to $\epsilon/L$) to Eq.~(\ref{sout}) can be obtained by matched asymptotic expansions \cite{match}. The inner expansion of $u\left(\mathbf{x}\right)$ is valid at distances from the escape hole that are much smaller than $L$. The outer expansion holds at distances much larger than $\epsilon$. The two expansions can then be matched in their joint region of validity to yield a composite expression valid across the entire domain. This method yields subleading corrections in $\epsilon/L$ for the non-interacting Brownian particles \cite{keller1,chevi1}. For interacting particles we can adopt a different
formalism. Remarkably, Eqs.~(\ref{ustatnd}) and (\ref{bqu1}) also serve as a simple model of thermal runaway in cooled chemical reactors, where $u\left(\mathbf{x}\right)$ is the stationary temperature field across a reactor that is insulated by its boundary except for a small cooling patch on it \cite{keller0,ward2}. The (important) difference is that in the NEP one also should evaluate the action and minimize it over possible multiple solutions.

The leading-order composite expression for $u\left(\mathbf{x}\right)$ coincides with Eq.~(\ref{u0}) \cite{keller0,ward2}. As we checked, the action (\ref{non-escape}) remains proportional to $f^2(n_0)$ up to, and including, the second order in $\epsilon/L$, with a geometry-dependent proportionality constant. The latter is given by the second-order expansion of the principle eigenvalue of the Laplace's operator $\mu_0^2$ \cite{second}. For a small absorbing disk of radius $\epsilon$ on the boundary of a sphere of radius $L$ one obtains $\mu_0^2 V=4\epsilon\left[1+\left(\epsilon/\pi L\right)\ln\left(\epsilon/L\right)+\dots \right]$ \cite{second}. In the context of the NEP of the SSEP, this leads to
\begin{equation}
s(n_0,\epsilon)\simeq 4D_0 \epsilon\left[1+\frac{\epsilon}{\pi L}\ln\left(\frac{\epsilon}{L}\right) \right] \arcsin^2(\sqrt{n_0}).\label{s2}
\end{equation}

Equations~(\ref{soutssep}) and~(\ref{s2}) hold for $D_0T\gg \epsilon^2$. However, they yield the MET of the first particle only for very low densities,  $n_0\epsilon^3 \ll 1$, where the inter-particle interactions can be neglected. For moderate and high densities, $n_0\epsilon^3 \gg 1$,
the MET of the first particle is much shorter than $\epsilon^2/D_0$. Here the optimal fluctuation for the non-escape is non-stationary, and we must return to the time-dependent MFT formulation~(\ref{path1}). The problem boils down to finding the survival probability $\mathcal P_{\text{1d}}$ of a gas of interacting particles, randomly distributed  on a half-line $x>0$,  against absorption at $x=0$. This problem was studied via the MFT \cite{MVK}. The stretched-exponential
decay with time, $-\ln {\mathcal P_{1d}} \simeq \sqrt{D(n_0)T}s_{\text{1d}}(n_0)$, holds in spite of the interactions. For the SSEP, the MFT yields a low-density expansion $s_{\text{1d}}(n_0) = (2/\sqrt{\pi})[n_0 + (\sqrt{2}-1) n_0^2 + \dots]$ \cite{MVK,Santos}. For higher densities $s_{\text{1d}}(n_0)$ can be computed numerically \cite{MVK}. This brings us to the result announced in Eq.~(\ref{shorttimeint}) with $g(n_0)\equiv s_{\text{1d}}(n_0)$, and we obtain $\langle T \rangle \simeq 2 [A_{\epsilon}^2 D_0 g^2(n_0)]^{-1}$.

A plausible setup, where our predictions can be compared to
experiment, is a ``pore-cavity-pore" device of $\mu$m dimensions with a nano-scale hole \cite{exp}. It allows for a controlled
entrapment of particles of a nano-scale size which, once trapped, can freely diffuse. Fluorescence imaging is used to track their positions. \blue{The authors of}
Ref. \cite{exp} reported measurements of the decay rate of the average number of particles inside the device, and noticed deviations from a
purely Brownian behavior. It would be interesting to also measure, for different initial number of particles, and different hole sizes,
the MET of the first particle from the device.

Finally, our general framework for the NEP, rooted in the MFT, can be extended to more complicated geometries \cite{bookz,geo1,geo2,geo3,geo4,geo5,geo6,chevi1,chevi2} and boundary conditions at the escape hole \cite{gleb}. It can also accommodate reactions among, and a finite lifetime of, the particles \cite{ElgartKamenev,reac1,reac2,hurtado2,reac3}.

\begin{acknowledgments}

We thank Gleb Oshanin and Zeev Schuss for useful discussions and acknowledge support from the Israel Science Foundation (Grant No. 807/16).

\end{acknowledgments}


\begin{thebibliography} {99}
\bibitem{keller1}
		M. J. Ward and J. B. Keller, SIAM J. Appl. Math. \textbf{53}, 770 (1993).
		
\bibitem{russians}
	I. V.  Grigoriev, Y. A. Makhnovskii, A. M. Berezhkovskii, and V. Y. Zitserman, J. Chem. Phys. \textbf{116}, 9574 (2002).
	
\bibitem{bress}
P. C. Bressloff and J. M. Newby, Rev. Mod. Phys. \textbf{85}, 135 (2013).
				
\bibitem{beni}
O. B\'{e}nichou and R. Voituriez, Phys. Rep. \textbf{539}, 225 (2014).

\bibitem{bookz}
D. Holcman and Z. Schuss, \textit{Stochastic Narrow Escape in Molecular and Cellular Biology}
(Springer, New York, 2015).

\bibitem{rev}
T. Chou and M. R. D'Orsogna, in ``\textit{First-Passage Phenomena and Their Applications}",
edited by R. Metzler,  G. Oshanin, and S. Redner (World Scientific, Singapore 2013).
	
\bibitem{diffapfere}
D. Coombs, R. Straube, and M. Ward, SIAM. J. Appl. Math. \textbf{70}, 302 (2009).

\bibitem{rna}
S. A. Gorski, M. Dundr, and T. Misteli, Curr. Opin. Cell Biol. \textbf{18}, 284 (2006).
	
	
\bibitem{den}
D. Holcman, Z. Schuss, and E. Korkotian, Bio. J. \textbf{87}, 81 (2004).
		

	

\bibitem{helmholtz} H. L. F. von Helmholtz, 
    J. Reine und Angewandte Mathematik \textbf{57}, 1 (1860).
	
\bibitem{rayleigh}
	J. W. S. Baron Rayleigh \textit{The Theory of Sound}, 2nd ed. (Dover, New York, 1945), Vol. 2.

\bibitem{beni2}
O. B\'{e}nichou and R. Voituriez, Phys. Rev. Lett. \textbf{100}, 168105 (2008).
	
	
\bibitem{geo1}	
D. Holcman, N. Hoze, and Z. Schuss, Phys. Rev. E \textbf{84}, 021906 (2011).
	
	\bibitem{geo2}	
E. Korkotian, D. Holcman, and M. Segal, Eur. J. Neurosci. \textbf{20},
2649 (2004).

	
\bibitem{geo3}
	Z. Schuss, \textit{Brownian Dynamics at Boundaries and Interfaces in Physics, Chemistry, and Biology} (Springer, New York, 2013).	
	
\bibitem{geo4}
	A. Singer, Z. Schuss, and D. Holcman,  J. Stat. Phys. \textbf{122},
	491 (2006).
	
\bibitem{geo5}
	D. Holcman and Z. Schuss, J. Phys. A \textbf{41}, 155001 (2008).
	
\bibitem{geo6}
	J. M. Arrieta, Trans. Amer. Math. Soc. \textbf{347} (1995).
	
\bibitem{chevi1}
	A. F. Cheviakov, M. J. Ward, and R. Straube, SIAM
	Multiscale Model. Simul. \textbf{8}, 836 (2010).
	
\bibitem{chevi2}
		A. F. Cheviakov and M. J. Ward, Math. Comput. Model. \textbf{53} (2011).

\bibitem{Greb2016} D. S. Grebenkov, Phys. Rev. Lett. \textbf{117}, 260201 (2016).

\bibitem{kill}
Z. Schuss, \textit{Theory and Applications of Stochastic Processes, An Analytical Approach}, Springer series on Applied Mathematical Sciences, Vol. 170, (Springer, New York, 2010).

\bibitem{kill2}
D. S. Grebenkov and J.-F. Rupprecht, J. Chem. Phys. \textbf{146}, 084106 (2017). 		


\bibitem{gleb}
D. S. Grebenkov and G. Oshanin, Phys. Chem. Chem. Phys. \textbf{19}, 2723 (2017).		

\bibitem{RoKim} S. Ro and Y. W. Kim, Phys. Rev. E \textbf{96}, 012143 (2017).

\bibitem{many2}
K. Basnayake, C. Guerrier, Z. Schuss, and D. Holcman, arXiv:1711.01330.

\bibitem{jackson}
J. D. Jackson,  \textit{Classical Electrodynamics} (Wiley, New York, 1999).


\bibitem{1d1}
M. Tachiya, Radiat. Phys. Chem. \textbf{21}, 167 (1983).

\bibitem{1d2} G. Zumofen, J. Klafter, and A. Blumen, J. Chem. Phys. \textbf{79}, 5131 (1983).

\bibitem{1d3}
 S. Redner and K. Kang, J. Phys. A Math. Gen. \textbf{17},  L451 (1984).


\bibitem{1d4}
 A. Blumen, G. Zumofen, and J. Klafter, Phys. Rev. B \textbf{30}, 5379(R) (1984).

\bibitem{1d5}
S. F. Burlatsky and A. A. Ovchinnikov, Sov. Phys. JETP \textbf{65}, 908 (1987).

\bibitem{1d6}
R. A. Blythe and A. J. Bray, Phys. Rev. E \textbf{67}, 041101  (2003).

\bibitem{1d7}
 J. Franke and S. N. Majumdar, J. Stat. Mech. P05024 (2012).

\bibitem{1d8}
 A. J. Bray, S. N. Majumdar, and G. Schehr, Adv. Phys. \textbf{62}, 225 (2013).

 \bibitem{1d9}
S. Redner and B. Meerson, J. Stat. Mech. P06019 (2014).

\bibitem{MVK}
B. Meerson, A. Vilenkin, and P. L. Krapivsky, Phys. Rev. E \textbf{90}, 022120 (2014).


\bibitem{Spohn}
H. Spohn, \textit{Large-Scale Dynamics of Interacting Particles} (Springer-Verlag, New York, 1991).

\bibitem{KL}
C. Kipnis and C. Landim, \textit{Scaling Limits of Interacting Particle Systems} (Springer, New York, 1999).


\bibitem{MFTreview}
	
L. Bertini, A. De Sole, D. Gabrielli, G. Jona-Lasinio, and C. Landim. Rev. Mod. Phys. \textbf{87}, 593 (2015).


	
\bibitem{MSR} P. C. Martin, E. D. Siggia, and H. A. Rose, Phys. Rev. A \textbf{8}, 423 (1973).

\bibitem{map}
J. Tailleur, J. Kurchan, and V. Lecomte, Phys. Rev. Lett. \textbf{99}, 150602 (2007); J. Tailleur, J. Kurchan, and V. Lecomte, J. Phys. A \textbf{41}, 505001 (2008).

\bibitem{deridamft}

B. Derrida and A. Gerschenfeld, J. Stat. Phys. \textbf{137}, 978 (2009).



\bibitem{Freid}
M.I. Freidlin and A.D. Wentzell, \textit{Random Perturbations of
	Dynamical Systems} (Springer-Verlag, New York, 1998).

\bibitem{surv}

T. Agranov, B. Meerson, and A. Vilenkin,
Phys. Rev. E \textbf{93}, 012136 (2016).

\bibitem{bd}

T. Bodineau and B. Derrida, Phys. Rev. Lett. \textbf{92}, 180601 (2004).


\bibitem{void}
	
P. L. Krapivsky, B. Meerson, and P. V. Sasorov, J. Stat. Mech. P12014 (2012).

	
\bibitem{conv}
Convergence of this integral
puts some limitations on the behavior of $D(\rho)$ and $\sigma(\rho)$ at small densities. As an example, let $D(\rho\rightarrow0)\sim\rho^{\alpha}$ and $\sigma(\rho\rightarrow0)\sim\rho^{\beta}$. Then the integral converges at $\rho\rightarrow0$ if and only if $2\alpha-\beta+2>0$. This condition holds in the examples we consider here.

\bibitem{boundary}
The condition $u(\mathbf{x}\in{\Omega_a})=0$ is inherited from
$\rho(\mathbf{x}\in{\Omega_a})=0$ due to the definition (\ref{transform1}). The condition $\nabla u(\mathbf{x}\in{\Omega_r})\cdot\hat{n}=0$ results from a boundary term that appears when minimizing the action (\ref{lang2}).

\bibitem{Liggett}
T. M. Liggett, \textit{Stochastic Interacting Systems: Contact, Voter,
			and Exclusion Processes} (Springer, New York, 1999).	


\bibitem{Paulbook} P. L. Krapivsky, S. Redner, and E. Ben-Naim, \textit{A Kinetic View of Statistical Physics} (Cambridge University Press, Cambridge, 2010).

\bibitem{match}
M. H. Holmes,  \textit{Introduction to Perturbation Methods} (Springer, New York, 1995).


\bibitem{keller0}
	
	M. J. Ward and J. B. Keller, Stud. Appl. Math. \textbf{85}, 1 (1991).
	
\bibitem{ward2}
	
	M. J. Ward and E. F. Van de Velde, J. Appl. Math. \textbf{48}, 53 (1992).

\bibitem{second}
	A. Singer, Z. Schuss, and D. Holcman, Phys. Rev. E \textbf{78},  051111 (2008).	

\bibitem{Santos} J. E. Santos and G. M. Sch\"{u}tz, Phys. Rev. E \textbf{64}, 036107 (2001).

\bibitem{exp}
D. Pedone, M. Langecker, G. Abstreiter, and U. Rant, Nano
	Lett. \textbf{11}, 1561 (2011).

\bibitem{ElgartKamenev}
V. Elgart and A. Kamenev, Phys. Rev. E \textbf{70}, 041106 (2004).

\bibitem{reac1}
T. Bodineau and M. Lagouge, J. Stat. Phys. \textbf{139}, 201 (2010).

\bibitem{reac2} B. Meerson and P. V. Sasorov, Phys. Rev. E \textbf{83}, 011129 (2011).

\bibitem{hurtado2} P. I. Hurtado, A. Lasanta, and A. Prados, Phys. Rev. E \textbf{88}, 022110 (2013).

\bibitem{reac3} B. Meerson, J. Stat. Mech. P05004 (2015).


\end{thebibliography}
\end{document}